\documentclass[12pt]{article}
\usepackage{amssymb}
\usepackage{epsfig}
\begin{document}
\begin{center}
{\large\bf NON-SINGULAR SPHERICALLY SYMMETRIC SOLUTION IN
EINSTEIN-SCALAR-TENSOR GRAVITY} \vskip 0.3 true in {\large J. W.
Moffat} \vskip 0.3 true in {\it The Perimeter Institute for
Theoretical Physics, Waterloo, Ontario, N2L 2Y5, Canada} \vskip
0.3 true in and \vskip 0.3 true in {\it Department of Physics,
University of Waterloo, Waterloo, Ontario N2L 3G1, Canada}
\end{center}
\begin{abstract}%

A static spherically symmetric metric in Einstein-scalar-tensor
gravity theory with a scalar field potential $V[\phi]$ is
non-singular for all real values of the coordinates. It does not
have a black hole event horizon and there is no essential
singularity at the origin of coordinates. The weak energy
condition $\rho_\phi > 0$ fails to be satisfied for $r\lesssim 1.3r_S$
(where $r_S$ is the Schwarzschild radius) but the strong energy condition $\rho_\phi+3p_\phi
> 0$ is satisfied. The classical Einstein-scalar-tensor solution is regular everywhere in spacetime without a black hole event horizon. However, the violation of the weak energy condition may signal the need for quantum physics anti-gravity as $r\rightarrow 0$. The non-singular static spherically symmetric solution is stable against the addition of ordinary matter.
\end{abstract}
\vskip 0.2 true in e-mail: john.moffat@utoronto.ca


\section{Introduction}

The problems associated with information loss of black holes have
been a source of controversy for more than thirty
years~\cite{Hawking}. The maximal extension of the Schwarzschild
spacetime by the Kruskal diagram~\cite{Kruskal}, completing the
space of geodesics except for the essential singularity at the
origin, has led to a general acceptance by the physics community
of the existence of black holes. There is evidence from the study
of the measured motions of stars in the close vicinity of Sgr A*
with a mass $M\sim 3.7\times 10^6\,M_\odot$ that a black hole
exists at the center of the Galaxy. However, due to the difficulty
of actually detecting a black hole event horizon as predicted by
general relativity (GR) the observational evidence remains
circumstantial and
controversial~\cite{Lasota,Narayan,Narayan2,Muller}.

A non-singular solution for cosmology~\cite{Moffat2} has been
obtained from a scalar-tensor-vector gravity (STVG)~\cite{Moffat}.
In the following, we shall give an example of a scalar-tensor
gravity theory which can yield a static spherically symmetric
solution which is free of an essential singularity and coordinate
event horizon.  A scalar field potential energy
allows for a more general spherically symmetric solution than
obtained in previous calculations in GR. The non-singular
solution of the field equations leads to a violation of the weak energy condition $\rho_\phi>0$ for $r\lesssim 1.3r_S$, where $r_S=2G_NM$ is the Schwarzschild radius, avoiding the Hawking-Penrose singularity theorem~\cite{Penrose, Hawking2,Ellis}. We attribute the violation of the weak energy condition for $r\lesssim 1.3r_S$ to the onset of quantum gravity or quantum repulsive exotic energy.

It is shown that when ordinary matter is added to the non-singular solution, then the absence of a singularity and an event horizon is maintained, for the scalar field barrier to singularity formation grows as the mass is added to the solution.

Because our non-singular spherically symmetric solution has
neither an event horizon nor a singularity at the origin, standard
black hole Hawking radiation is absent for a static, spherically symmetric astrophysical object and there is no Hawking information loss paradox.
The simpler Einstein scalar-tensor gravity theory with a scalar
field potential that can yield a non-singular static spherically
symmetric solution can be generalized to a non-singular solution
of the STVG field equations~\cite{Moffat}.

\section{The Action and the Field Equations}

The action takes the form:
\begin{equation}
S=S_{\rm Grav}+S_\phi+S_M,
\end{equation}
where
\begin{equation}
S_{\rm Grav}=\frac{1}{16\pi G_N}\int d^4x\sqrt{-g}(R+2\Lambda),
\end{equation}
\begin{equation}
S_\phi=-\int
d^4x\sqrt{-g}\biggl[\frac{1}{2}\partial^\mu\phi\partial_\mu\phi-V[\phi]\biggr].
\end{equation}
Here, $R$ is the Ricci scalar $R=g^{\mu\nu}R_{\mu\nu}$, $\Lambda$
is Einstein's cosmological constant, $\phi$ is a scalar field and
$V[\phi(r)]$ denotes a $\phi$ field potential energy. We use the
metric signature $\eta_{\mu\nu}={\rm diag}(1,-1,-1,-1)$ where
$\eta_{\mu\nu}$ denotes the Minkowski metric tensor and we choose
(unless otherwise stated) units with the speed of light $c=1$.

We have
\begin{equation}
T_{\mu\nu}=-\frac{2}{\sqrt{-g}}\frac{\delta S_M}{\delta
g^{\mu\nu}},
\end{equation}
where $T_{\mu\nu}$ is the total stress-energy momentum tensor
\begin{equation}
T_{\mu\nu}=T_{M\mu\nu}+T_{\phi\mu\nu},
\end{equation}
and $T_{M\mu\nu}$ and $T_{\phi\mu\nu}$ denote the matter and
scalar field energy-momentum tensors, respectively. We have
\begin{equation}
\label{Tphi} T_{\phi\mu\nu}=\partial_\mu\phi\partial_\nu\phi
-g_{\mu\nu}\biggl(\frac{1}{2}\partial^\alpha\phi\partial_\alpha\phi-V[\phi]\biggr).
\end{equation}

The gravitational field equations are given by
\begin{equation}
\label{Einsteineqs}G_{\mu\nu}-g_{\mu\nu}\Lambda=8\pi
G_NT_{\mu\nu},
\end{equation}
where $G_{\mu\nu}=R_{\mu\nu}-\frac{1}{2}g_{\mu\nu}R$ is the Einstein tensor. From the Bianchi identities $\nabla_\nu G^{\mu\nu}=0$ we obtain the
conservation law:
\begin{equation}
\nabla_\nu T^{\mu\nu}=0,
\end{equation}
where $\nabla_\mu$ denotes the covariant derivative with respect
to the metric $g_{\mu\nu}$. The field $\phi$ satisfies the
equation of motion
\begin{equation}
\label{phiequation} \nabla^\mu\nabla_\mu\phi+\frac{\partial
V[\phi]}{\partial\phi}=0.
\end{equation}

\section{Non-Singular Static Spherically Symmetric Solution}

The line element is of the standard form for a time-dependent spherically
symmetric metric:
\begin{equation}
\label{staticmetric} ds^2=B(r,t)dt^2-A(r,t)dr^2-r^2d\Omega^2,
\end{equation}
where
\begin{equation}
d\Omega^2=d\theta^2+\sin^2\theta d\phi^2.
\end{equation}
The field equations for $T_{M\mu\nu}=0$ and $\Lambda=0$ are given by
\begin{equation}
\label{Riccieq} R_{\mu\nu}=8\pi
G_N(T_{\phi\mu\nu}-\frac{1}{2}g_{\mu\nu}T_\phi),
\end{equation}
where $T_\phi=g^{\mu\nu}T_{\phi\mu\nu}$. We have
\begin{equation}
T_{\phi\mu\nu}-\frac{1}{2}g_{\mu\nu}T_\phi=\partial_\mu\phi\partial_\nu\phi
-g_{\mu\nu}V[\phi].
\end{equation}
The Ricci tensor is given by
\begin{eqnarray}
R_{00}&=&\frac{B''-\ddot{A}}{2A}+\frac{\dot{A}\dot{B}-B'^2}{4AB}+\frac{\dot{A}^2-A'B'}{4A^2}+\frac{B'}{Ar},\\
R_{rr}&=&\frac{\ddot{A}-B''}{2B}+\frac{B'^2-\dot{A}\dot{B}}{4B^2}+\frac{A'B'-\dot{A}^2}{4AB}+\frac{A'}{Ar},\\
\label{r0eq}\nonumber\\
R_{r0}=R_{0r}&=&\frac{\dot{A}}{Ar}\\
R_{\theta\theta}=\frac{1}{\sin^2{\theta}}R_{\phi\phi}&=&1-\frac{1}{A}+\frac{A'r}{2A^2}-\frac{B'r}{2AB}.
\end{eqnarray}

We have for the right-hand side of (\ref{Riccieq}):
\begin{eqnarray}\\
T_{\phi\mu\nu}&-&\frac{1}{2}g_{\mu\nu}T_\phi\nonumber\\
&=&\left(\begin{array}{cccc}
\dot{\phi}^2-g_{00}V[\phi]&\dot{\phi}\phi'&0&0\\
\dot{\phi}\phi'&\phi'^2-g_{rr}V[\phi]&0&0\\
0&0&-g_{\theta\theta}V[\phi]&0\\
0&0&0&-g_{\phi\phi}V[\phi]
\end{array}\right).
\label{Ttensor}
\end{eqnarray}

In the static case ${\dot A}={\dot B}=0$ and $\dot\phi=0$, we obtain
\begin{eqnarray}
R_{00}&=&-8\pi G_Ng_{00}V[\phi(r)],\\
R_{\theta\theta}&=&-8\pi G_Ng_{\theta\theta}V[\phi(r)],
\end{eqnarray}
or
\begin{equation}
\frac{R_{00}}{R_{\theta\theta}}=\frac{g_{00}}{g_{\theta\theta}}.
\label{Ratioeq}
\end{equation}
This equation can be shown to give
\begin{equation}
\label{Bdivr}
\frac{2ABB''r-AB'^2r-A'BB'r+4ABB'}{4A^2B-4AB+2A'Br-2AB'r}=-\frac{B}{r}.
\end{equation}
A useful relationship between $A(r)$ and $B(r)$ can be derived:
\begin{equation}
\label{Aeq}
4(A-1)+\left(2\frac{A'}{A}+2\frac{B'}{B}\right)r+\left(2\frac{B''}{B}-\frac{B'^2}{B^2}
-\frac{A'}{A}\frac{B'}{B}\right)r^2=0.
\end{equation}

Let us take
\begin{equation}
\label{BA}
B(r)=1/A(r).
\end{equation}
We obtain from Eq.(\ref{Aeq}) the solution
\begin{equation}
B(r)=1+\frac{C_1}{r}+C_2r^2.
\end{equation}
Choosing $C_1=-r_S=-2G_NM$ and $C_2=0$, we arrive at the Schwarzschild metric:
\begin{equation}
ds^2=\biggl(1-\frac{r_S}{r}\biggr)dt^2-\frac{1}{1-\frac{r_S}{r}}dr^2-r^2d\Omega^2.
\end{equation}
We note that the pressure terms in the energy-momentum tensor $T_{\phi\mu\nu}$ (\ref{Ttensor}) for the Einstein-scalar theory of gravity are not isotropic.

Let us now consider the case when
\begin{equation}
\label{BAnot}
B(r)\not=1/A(r).
\end{equation}
We see that even though ${\dot A}=0$ from Eqs.(\ref{r0eq}) and (\ref{Ttensor}), we still have ${\dot B}\not=0$. Therefore, when (\ref{BAnot}) holds the solution for the metric does not satisfy the Birkhoff theorem ~\cite{Birkhoff}  as in the case when (\ref{BA}) is satisfied for the Schwarzschild solution.

Using $\alpha=\ln{A}$ and $\beta=\ln{B}$ such that $A'/A=\alpha'$,
$B'/B=\beta'$, and $B''/B=\beta''+\beta'^2$, we get
\begin{equation}
4(\exp(\alpha)-1)+\left(2r-\beta'r^2\right)\alpha'+2\beta'r+\left(2\beta''+\beta'^2\right)r^2=0.
\end{equation}
This equation is solvable for $\alpha$. We have
\begin{equation}
f(r)\alpha'(r)+\exp(\alpha(r))+g(r)=0,
\end{equation}
from which
\begin{equation}
\alpha(r)=-\int
dr{\frac{g(r)}{f(r)}}-\ln\left\{C+\int dr{\frac{\exp{\left[-\int
dr{\frac{g(r)}{f(r)}}\right]}}{f(r)}}\right\},
\end{equation}
where $C$ is a constant. Given
\begin{eqnarray}
f(r)&=&\frac{1}{2}r-\frac{1}{4}\beta'r^2,\\
g(r)&=&\frac{1}{2}\beta'r+\frac{1}{2}\beta''r^2+\frac{1}{4}\beta'^2r^2-1,
\end{eqnarray}
we get
\begin{equation}
\label{alphaeq} \alpha(r)=\int
dr{\frac{4-2\beta'r-2\beta''r^2-\beta'^2r^2}{r(2-\beta'r)}}
-\ln\left\{C+4\int
dr{\frac{\exp{\left[\int dr{\frac{4-2\beta'r-2\beta''r^2-\beta'^2r^2}{r(2-\beta'r)}}
\right]}}{r(2-\beta'r)}}\right\}.
\end{equation}
Further,
\begin{equation}
\alpha'=\beta'+\frac{2}{r}-2\frac{\beta''r+\beta'}{2-\beta'r}-\frac{e^\beta(2r-\beta'r^2)}{\frac{C}{4}
+\int dr{e^\beta(2r-\beta'r^2)}}.
\end{equation}
Therefore,
\begin{equation}
\label{A2eq}
A(r)=\frac{(2B(r)r-B'(r)r^2)^2}{B(r)(C+8\int dr{(B(r)r-B'(r)r^2))}}.
\label{eq:19}
\end{equation}

Let us consider the {\it ansatz} for $g_{00}(r)= B(r)$:
\begin{equation}
\label{Bequation}
B(r)=(1/a)[a-1+\exp{(-ar_S/r)}].
\end{equation}
We see that $B(r)$ does not have an event horizon for any real value of $r$
in the range $0\leq r\leq \infty$ for $a>1$. Moreover, $B(r)$ is
non-singular at $r=0$:
\begin{equation}
B(0)=\frac{a-1}{a}.
\end{equation}
Moreover, we have
\begin{equation}
B(r)\sim 1-\frac{r_S}{r},
\end{equation}
so that $B(r)$ satisfies the Schwarzschild solution for large $r$, and the metric line element satisfies the Minkowski spacetime boundary condition as $r\rightarrow \infty$.

We shall use
\begin{equation}
\beta(r)=\ln{B(r)}=\ln{\left\{(1/a)[a-1+\exp{(-ar_S/r)}]\right\}},
\end{equation}
to obtain from Eq.(\ref{A2eq}) the solution
\begin{equation}
\label{Aequation}
A(r)=\frac{\left[2(a-1)r+(2r-ar_S)\exp{\left(-\frac{ar_S}{r}\right)}\right]^2}
{aB(r)\left\{C+4ar^2B(r)+8ar_S\left[ar_SE_1\left(\frac{ar_s}{r}\right)
-r\exp{\left(-\frac{ar_S}{r}\right)}\right]\right\}},
\end{equation}
with $E_1$ denoting the exponential integral of the first kind.

If $a=2$ and the integration constant $C=0$, we get
\begin{equation}
\lim_{r\rightarrow+0}A(r)=1.
\end{equation}
To first order, $A(r)$ evaluates to
\begin{equation}
A(r)\simeq \frac{1}{1-\frac{r_S}{r}},
\end{equation}
and our metric satisfies the Schwarzschild solution for large values
of $r$. Plots of the metric components $A(r)$ and $B(r)$ are shown in
Fig.1. We observe that $A(r)$ and $B(r)$ are non-singular for all real
values in the range $0\leq r \leq \infty$. Because the lowest order behavior of $B(r)$ and $A(r)$ for large values of $r$ is the same as the Schwarzschild solution and the higher order contributions are small, then our solution agrees with all the classical gravitational experiments in the solar system and the binary pulsar observations.

\begin{figure}
\begin{center}
\begin{minipage}[b]{0.5\linewidth}
{\psfig{file=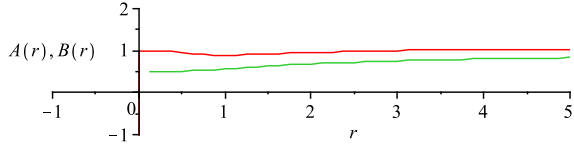,width=\linewidth}}
\end{minipage}
\end{center}
\caption{Plots of $A(r)$ (red curve) and $B(r)$ (green curve) versus $r$ for $a=2$.}
\end{figure}

By using the components of the Einstein tensor:
\begin{equation}
\label{G00}
{G_0}^0=\frac{1}{rA}\biggl(\frac{A'}{A}-\frac{1}{r}\biggr)+\frac{1}{r^2},\quad {G_r}^r=-\frac{1}{rA}\biggl(\frac{B'}{B}+\frac{1}{r}\biggr)+\frac{1}{r^2},
\end{equation}
we can evaluate $\rho_\phi$ and $\rho_\phi+3p_\phi$ from the field equations
and the energy-momentum tensor (\ref{Ttensor}). We find that the non-singular solution violates the weak energy condition $\rho_\phi
\geq 0$ for $r\lesssim 1.3 r_s$ but not the strong energy condition
$\rho_\phi+3p_\phi \geq 0$. Plots of $\rho_\phi$ and $\rho_\phi+3p_\phi$ are shown in Fig. 2.

\begin{figure}
\begin{center}
\begin{minipage}[b]{0.5\linewidth}
{\psfig{file=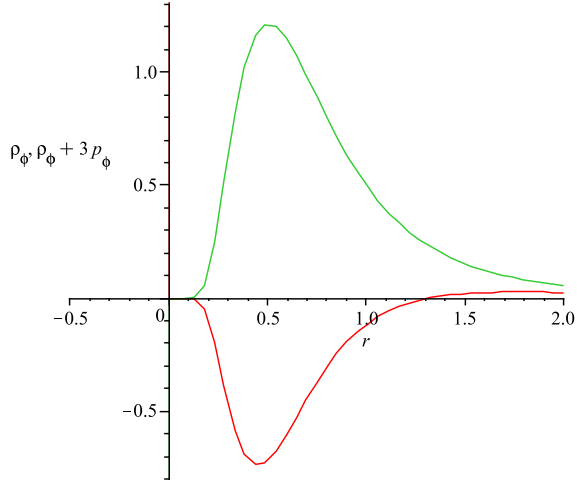,width=\linewidth}}
\end{minipage}
\end{center}
\caption{Plots of $\rho_\phi(r)$ (red curve) and $\rho_\phi(r)+3p_\phi(r)$ (green curve) versus $r$.}
\end{figure}

The Ricci scalar $R$ and the Kretschmann curvature invariant given by
\begin{equation}
K=R^{\mu\nu\rho\sigma}R_{\mu\nu\rho\sigma}
\end{equation}
also remain well-behaved at $r=0$ as shown in Figs. 3 and 4.

\begin{figure}
\begin{center}
\begin{minipage}[b]{0.5\linewidth}
{\psfig{file=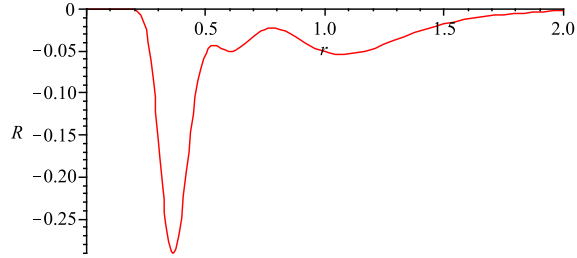,width=\linewidth}}
\end{minipage}
\end{center}
\caption{This displays the Ricci scalar curvature invariant $R$.}
\end{figure}

\begin{figure}
\begin{center}
\begin{minipage}[b]{0.5\linewidth}
{\psfig{file=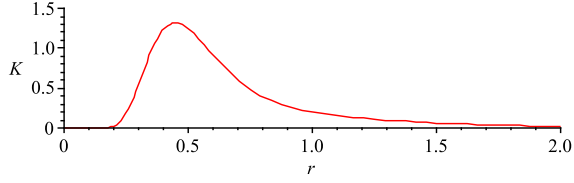,width=\linewidth}}
\end{minipage}
\end{center}
\caption{This displays the Kretschmann curvature invariant $K$.}
\end{figure}

The total mass of the static spherically symmetric solution is
calculated from the following integral:
\begin{equation}
M=\frac{1}{2G_N}\int_0^\infty drr^2{G_0}^0(r),
\end{equation}
where the Einstein tensor components ${G_0}^0$ are given by (\ref{G00}).
This integral is quite complicated and cannot be evaluated in
closed form. However, when evaluated numerically using $C=0$ and
$a\ge 3/2$ for arbitrary values of $r_S$, we get
\begin{equation}
M=\frac{1}{2G_N}r_S.
\end{equation}

The geodesic equations for a test particle are given by
\begin{eqnarray}
0&=&\frac{B'}{B}\frac{dr}{dt}\frac{dr}{ds}+\frac{d^2t}{ds^2},\\
0&=&\frac{B'}{2A}\left(\frac{dt}{ds}\right)^2
+\frac{A'}{2A}\left(\frac{dr}{ds}\right)^2+\frac{d^2r}{ds^2}
-\frac{r}{A}\left[\left(\frac{d\theta}{ds}\right)^2+\left(\frac{d\phi}{ds}\right)^2\sin^2\theta\right],\\
0&=&\frac{d^2\theta}{ds^2}
+\frac{2}{r}\frac{dr}{ds}\frac{d\theta}{ds}
-\left(\frac{d\phi}{ds}\right)^2\sin\theta\cos\theta,\\
0&=&\frac{d^2\phi}{ds^2}
+\frac{2}{r}\frac{dr}{ds}\frac{d\phi}{ds}+2\frac{d\theta}{ds}\frac{d\phi}{ds}\cot\theta.
\end{eqnarray}
We have for $a=2$: $A(0)=1$, $B(0)=1/2$, $A'(0)=0$ and $B'(0)=0$,
and the above equations reduce to the geodesic equations:
\begin{eqnarray}
0&=&\frac{d^2t}{ds^2},\\
0&=&\frac{d^2r}{ds^2}-r\left[\left(\frac{d\theta}{ds}\right)^2
+\left(\frac{d\phi}{ds}\right)^2\sin^2\theta\right],\\
0&=&\frac{d^2\theta}{ds^2}+\frac{2}{r}\frac{dr}{ds}\frac{d\theta}{ds}
-\left(\frac{d\phi}{ds}\right)^2\sin\theta\cos\theta,\\
0&=&\frac{d^2\phi}{ds^2} +\frac{2}{r}\frac{dr}{ds}\frac{d\phi}{ds}
+2\frac{d\theta}{ds}\frac{d\phi}{ds}\cot\theta.
\end{eqnarray}
for the Minkowski metric in spherical polar coordinates:
\begin{equation}
ds^2=dt^2-dr^2-r^2d\Omega^2.
\end{equation}
In other words, at $r=0$ the geodesic equations are the vacuum
geodesic equations. The test particle equations for the non-singular metric components $B(r)$ and $A(r)$ are geodesically complete.

Let us now consider the $V[\phi(r)]$ that can be obtained from the
field equations:
\begin{equation}
R_{\theta\theta}=-8\pi G_N g_{\theta\theta}V[\phi],
\end{equation}
or
\begin{equation}
V[\phi(r)]=-\frac{R_{\theta\theta}}{8\pi G_Nr^2}.
\end{equation}
We have
\begin{equation}
R_{rr}=8\pi G_N\left(\phi'^2-g_{rr}V[\phi]\right)=8\pi
G_N\left(\phi'^2+A\frac{R_{\theta\theta}}{8\pi G_Nr^2}\right),
\end{equation}
from which we obtain
\begin{eqnarray}
\phi'&=&\sqrt{\frac{1}{8\pi G_N}\left(R_{rr}-\frac{A}{r^2}R_{\theta\theta}\right)}\\
&=&\sqrt{\frac{1}{8\pi G_N}\left(\frac{-B''}{2B}+\frac{B'^2}{4B^2}+\frac{A'B'}{4AB}
+\frac{A'}{Ar}-\frac{A}{r^2}+\frac{1}{r^2}-\frac{A'}{2Ar}+\frac{B'}{2Br}\right)}\\
&=&\sqrt{\frac{-1}{32\pi
G_Nr^2}\left[4(A-1)-\left(2\frac{A'}{A}+2\frac{B'}{B}\right)r+\left(2\frac{B''}{B}
-\frac{B'^2}{B^2}-\frac{A'B'}{AB}\right)r^2\right]}.
\end{eqnarray}
Comparing this result with Eq.~(\ref{Aeq}) allows us to eliminate
many terms, leaving us with
\begin{equation}
\phi'=\sqrt{\frac{1}{8\pi G_N
r^2}\left(\frac{A'}{A}+\frac{B'}{B}\right)r}=\sqrt{\frac{(\ln{AB})'}{8\pi
G_Nr}}.
\end{equation}
From this, it can be seen that unless the product $AB$ is
monotonically increasing, $\phi'$ becomes imaginary. We can also
use Eq.~(\ref{A2eq}) to obtain
\begin{equation}
\phi'=\sqrt{\frac{\left(\ln{\frac{(2Br-B'r^2)^2}{C+8\int dr{(Br-B'r^2)}}}\right)'}{8\pi
G_Nr}}.
\end{equation}
In the case of $B(r)=(1/a)[a-1+\exp(-ar_S/r)]$, we get the
explicit result,
\begin{eqnarray}
\phi'=&2ar_S&\left\{\left[\left(r^2+ar_Sr-\frac{1}{2}a^2r_S^2\right)E_1\left(\frac{ar_S}{r}\right)
-\frac{(a-1)r^2}{4}\right]\exp{\left(\frac{-ar_S}{r}\right)}\right.\nonumber\\
&&+\left.\left[\left(\frac{ar_S}{2}
-\frac{11r}{8}\right)\exp{\left(\frac{-2ar_S}{r}\right)
+(a-1)rE_1\left(\frac{ar_S}{r}\right)}\right]r\right\}^{1/2}\nonumber\\
&\times&\left\{\left[r(r-2ar_S)\exp{\left(\frac{-ar_s}{r}\right)}
+2a^2r_S^2E_1\left(\frac{ar_S}{r}\right)
+(a-1)r^2\right]\right.\nonumber\\
&&\times\left.\left[\left(r-\frac{ar_S}{2}\right)\exp{\left(\frac{-ar_S}{r}\right)}
+(a-1)r\right]r^3\right\}^{-1/2}.
\end{eqnarray}
In Fig. 4, we plot $\phi^{'2}$ versus $r$ for $a=2$ and $r_S=1$.

\begin{figure}
\begin{center}
\begin{minipage}[b]{0.5\linewidth}
{\psfig{file=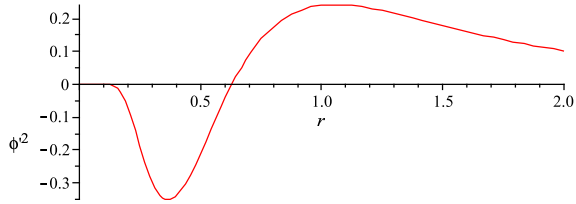,width=\linewidth}}
\end{minipage}
\end{center}
\caption{Plot of $\phi^{'2}$ versus $r$.}
\end{figure}

We also calculate $V[\phi(r)]$:
\begin{eqnarray}
V[\phi(r)]=&a^2r_S^2&\left\{\left(\frac{r}{2}
+\frac{ar_S}{8}\right)r\exp\left(\frac{-2ar_S}{r}\right)\right.\nonumber\\
&&\left.+\left[(a-1)(ar_S-r)
-ar_SE_1\left(\frac{ar_S}{r}\right)\right]r\exp\left(\frac{-ar_S}{r}\right)\right.\nonumber\\
&&\left.+(a-1)\left[ar_S(ar_S-r)E_1\left(\frac{ar_S}{r}\right)
+\frac{a-1}{2}r^2\right]\right\}\nonumber\\
&\times&\left\{\pi G_N\left[(ar_S-2r)\exp{\left(\frac{-ar_S}{r}\right)
-2(a-1)r}\right]^3r^3\exp\left(\frac{ar_S}{r}\right)\right\}^{-1}.
\end{eqnarray}
A plot of $V[\phi(r)]$ with $a=2$ and $r_S=1$ is shown in
Fig.5:

\begin{figure}
\begin{center}
\begin{minipage}[b]{0.5\linewidth}
{\psfig{file=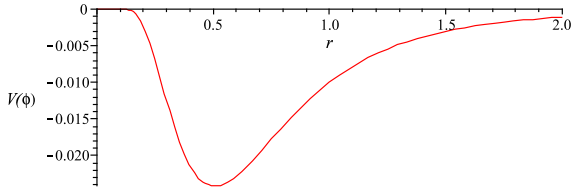,width=\linewidth}}
\end{minipage}
\end{center}
\caption{Plot of $V[\phi(r)]$ versus $r$.}
\end{figure}

We observe that $\phi^{'}(r)$ becomes imaginary for $r < r_S$, while $V[\phi(r)]$ is negative corresponding to a
repulsive potential and approaches zero as $r\rightarrow\infty$. The imaginary behavior of $\phi'(r)$ for $r < r_S$
means that a quantum behavior of the scalar field energy is required to explain the exotic form of the scalar field density $\rho_\phi$ needed to remove
a singularity at $r=0$. This could be quantum gravity or some form of quantum repulsive scalar field energy. On the other hand, all the physical properties of our regular solution, such as $\rho_\phi$ and $V[\phi(r)]$ are bounded, so this may mean that we do not require quantum gravity to obtain non-singular solutions in gravity theory.

In the range $0\leq r \leq \infty$, the strong energy condition
$\rho_\phi+3p_\phi \geq 0$ or $R_{\mu\nu}U^\mu U^\nu \geq 0$ for a
null or timelike vector $U^\mu$ is not violated. However, the weak
energy condition: $\rho_\phi \geq 0$ is violated as we approach the
Schwarzschild radius $r_S$. The Hawking-Penrose
theorems~\cite{Penrose,Hawking2,Ellis} state that if both the weak
and strong energy conditions for matter are satisfied and there
exists an apparent event horizon, then the static spherically
symmetric solution in GR must be singular. In particular, the
focussing of null and time-like geodesics produces an event
horizon at $r_S$ and an essential singularity at $r=0$. Since our
non-singular static spherically symmetric metric violates the weak
energy condition for $0\leq r\lesssim 1.3r_S$, we do not contradict the
Hawking-Penrose theorems. Because of the absence of an event
horizon {\it our exterior solution does not possess a trapped
surface} and the geodesics of the spacetime metric are
complete~\cite{Penrose,Hawking2,Ellis}.

The violation of the weak energy condition signals that the {\it
classical} non-singular solution fails for $r \lesssim 1.3r_S$. As
$r\rightarrow 0$ quantum physics must intervene to preserve the
non-singular solution and prevent the existence of an event
horizon and an essential singularity at $r=0$.

If we add ordinary matter density $\rho_M$ to the right-hand side of our field equations:
\begin{equation}
\rho=\rho_M+\rho_\phi,
\end{equation}
then we observe that the non-singular metric solutions $B(r)$ and $A(r)$ and the potential $V[\phi(r))]$ only depend on M through the Schwarzschild radius $r_S=2G_NM$. Therefore, as mass $\delta M$ is added:
\begin{equation}
M=M+\delta M
\end{equation}
the size of the repulsive potential barrier and the negative contribution of $\rho_\phi$ increase and continue to maintain a stable non-singular metric. Thus, the scalar field acts as an {\it effective, quantum exotic matter-energy} that prevents the existence of an essential singularity in the metric at $r=0$ and the formation of a black hole event horizon.

\section{Conclusions}

We have proposed an exterior non-singular static spherically
symmetric solution for Einstein-scalar field gravity. The scalar
field allows for negative density $\rho_\phi$ and the potential $V[\phi(r)]$ is repulsive as a function of $r$ and $M$.

When the nuclear fuel burns out in cores of stars, then in
standard GR the star can collapse to zero radius and form an event
horizon with an essential singularity at $r=0$. The Fermi pressure
at the core of a collapsed star caused by the degenerate electron
and neutron gas and Pauli's exclusion principle stabilizes white
dwarfs and neutron stars, respectively. For the stable white
dwarfs and neutron stars the Chandrasekhar mass limits are:
$M_c\sim 1.4\, M_{\odot}$ and $M_c\sim 2-3\, M_{\odot}$,
respectively~\cite{Ohanian}. When the mass of the star satisfies $M > 6\, M_{\odot}$, then according to the singular solutions of GR the star collapses to a black hole. However, for our non-singular solution, even when the star has a mass $M > 6\, M_{\odot}$, the effective quantum field density $\rho_\phi < 0$ as $r\rightarrow 0$ and the
repulsive potential $V[\phi(r)]$ prevents the star from forming a singularity at $r=0$ and a black hole event horizon. At the core of
the collapsed star the density of negative field energy $\rho_\phi$
can be sufficiently large to prevent the star from collapsing to a
black hole as the mass $M$ is increased. Thus, if an effective matter-energy $\rho_\phi$ dominates as $r\rightarrow 0$, then we can expect that at the core of a compact star the energy density $\rho_\phi$ can produce enough ``antigravity'' to form a stable ``dark grey'' or ``black'' star, even as we continue to increase the ordinary matter $M$ for the star. An important issue justifying further investigation is whether a non-singular collapsed dark grey star is stable. The non-singular solution must also be extended to the case of a rotating grey or black star.

We can speculate that the scalar $\phi$ field energy permeates all of spacetime as a form of ``quintessence''. This would mean that the vacuum field equations
in GR:
\begin{equation}
R_{\mu\nu}=0
\end{equation}
are never fulfilled and that if our potential $V[\phi(r)]$ is valid for a collapsed star, then black holes as described by the Schwarzschild solution do not exist in nature.

Our classical solution to Einstein-scalar-tensor gravity is non-singular throughout spacetime with bounded curvature invariant $K$, $\rho_\phi$, $p_\phi$, and $V[\phi(r)]$.  On the other hand, the fact that $\rho_\phi$ becomes negative as $r\rightarrow 0$ could signal the need for quantum gravity at short distances to yield a solution that is regular everywhere in spacetime.

Because our static spherically symmetric solution does not possess
a black hole event horizon, the compact star will only
radiate ``normal'' radiation. This radiation may be so small that
the star appears to an outside observer to be ``black''.
Since Hawking radiation is intimately associated with a black hole
event horizon in the Schwarzschild solution of GR and such an
event horizon is absent in our exterior regular solution, then our
dark grey star does not have an information loss problem.

\vskip 0.2 true in {\bf Acknowledgments} \vskip 0.2 true in

This work was supported by the Natural Sciences and Engineering
Research Council of Canada. Research at the Perimeter Institute
for Theoretical Physics is supported by the Government of Canada
through NSERC and by the Province of Ontario through the Ministry
of Research and Innovation (MRI). I thank Martin Green, Robert
Mann and Parampreet Singh for helpful and stimulating discussions. I thank Viktor Toth for valuable assistance in calculating equations using a Maxima
tensor package~\cite{Toth} and Maple and for many helpful discussions.

\end{document}